\documentclass[twocolumn,aps,floatfix]{revtex4}
\usepackage{amssymb}
\usepackage{graphicx}

\begin{document}
 
\title{Dynamics of quasisolitons in degenerate fermionic gases}
 
\author{Emilia Witkowska$\,^1$ and Miros{\l}aw Brewczyk$\,^2$}
 
\affiliation{\mbox{$^1$Instytut Fizyki  PAN, Aleja Lotnik\'ow 32/46, 
                       02-668 Warsaw, Poland}  \\
\mbox{$^2$Instytut Fizyki Teoretycznej, Uniwersytet w Bia{\l}ymstoku, 
                ulica Lipowa 41, 15-424 Bia{\l}ystok, Poland}  \\  }
\date{\today} 

\begin{abstract}
We investigate the dynamics of the system of multiple bright and dark 
quasisolitons generated in a one-component ultracold Fermi gas via the 
phase imprinting technique in terms of atomic orbitals approach. 
In particular, we analyze the collision between two bright quasisolitons 
and find that quasisolitons are subject to the superposition principle.

\end{abstract}

\maketitle

Solitons are already well established phenomena in the area of dilute atomic
quantum gases. They have been observed in several experiments with trapped
bosonic atoms as well as described in many theoretical papers. The latter
usually involves the mean-field approach and the resulting Gross-Pitaevskii
equation for the condensate wave function. The intrinsic nonlinearity of
this equation is originating from the interparticle interactions. As the
effective interaction between atoms can be both repulsive or attractive,
two types of solitonic excitations \cite{Zakharov} are possible in
Bose-Einstein condensates. 

Dark solitons were realized experimentally first \cite{dark1,dark2,dark3} 
whereas there have been problems with generating bright solitons implied 
by the occurrence of the collapse in the attractive condensates.
Finally, two ways turned out to be successful. In experiments of Refs. 
\cite{bright1,bright2} a large condensate with positive scattering length
is prepared and next the interatomic interactions are changed to attractive
ones (with small negative value of the scattering length) by using
Feshbach resonances. In this case the single bright soliton or a train
of solitons were observed. Another way of creating bright solitons was
demonstrated in the experiment of Ref. \cite{negmass} where the condensate
still remains repulsive but appropriate engineering of atoms in a periodic
potential allows for a change of the sign of the effective mass and results
in the generation of bright gap solitons. Recently, another concept of bright 
matter waves in a repulsive condensate has been developed that involves 
degenerate Fermi gas as a ''stabilizing medium" \cite{Tomek}.

In a series of papers \cite{Tomek1,Tomek2} we have shown that similar
structures can also be created in a one-component ultracold fermionic
gases even though to a very good approximation spin-polarized fermions
at zero temperature do not interact. These structures exist both in the
systems with uniform and nonuniform densities under a quasi-one-dimensional
confinement. They can be generated by using the technique of phase imprinting
and its appearance is related to the Fermi statistics. There exists the
regime of parameters, defined by the condition that the width of the imprinted 
phase is bigger than the Fermi length, where only states with momentum close 
to the Fermi momentum are excited. Hence, the life time of generated structures 
can be long enough (in comparison with the Fermi time) allowing for observation
and making them similar to solitons.

The method of phase imprinting has been already employed in the experiments
where dark solitons have been generated \cite{dark1,dark2}. The strong and
short enough off-resonance laser pulse is passing first through the
appropriately tailored absorption plate and next through the sample of
atoms. Under such conditions the motion of atoms is frozen and the only
result of the interaction of light with the atoms is imprinting the phase
on the atomic wave functions. After the light is gone, the density and
the phase change and desired structures (like solitons or vortices) appear 
in the system, according to the pattern written in the absorption plate. 

In this paper we further investigate the motion of fermionic quasisolitons
by analytical calculations based on the propagator technique. These
calculations allow us for a detailed description of the collision between
fermionic quasisolitons. We come about the conclusion that quasisolitons
fulfill the superposition principle, i.e. the sum of two quasisolitons is
a two-quasisoliton solution.

So, we consider a system of $N$ fermionic atoms confined in a one-dimensional
box with periodic boundary conditions at zero temperature. We assume that 
the many-body wave function of such a system is well described by the Slater 
determinant

\begin{eqnarray}
\Psi (x_1,...,x_N) = \frac{1}{\sqrt{N!}} \left |
\begin{array}{lllll}
\varphi_1(x_1) & . & . & . & \varphi_1(x_N) \\
\phantom{aa}. &  &  &  & \phantom{aa}. \\
\phantom{aa}. &  &  &  & \phantom{aa}. \\
\phantom{aa}. &  &  &  & \phantom{aa}. \\
\varphi_N(x_1) & . & . & . & \varphi_N(x_N)
\end{array}
\right | \, . 
\label{Slater}
\end{eqnarray}
The gas is spin-polarized and therefore the s-wave scattering, dominant at
low temperatures, is absent and effectively the gas becomes the system
of noninteracting particles. The many-body Schr\"odinger equation is then
equivalent to the set of one-particle equations, each of them describing the
evolution of a particular single-particle orbital $\varphi_1(x),...,\varphi_N(x)$
under the same Hamiltonian. Thus these orbitals remain orthogonal during the 
evolution and the one-particle density matrix is at any time given by the 
following formula
\begin{equation}
\rho_1(x^{\, \prime}, x^{\, \prime \prime}, t) = \frac{1}{N}
\sum_{n=1}^N \varphi_n(x^{\, \prime}, t) \; 
\varphi_n^*(x^{\,\prime \prime}, t) \, ,
\end{equation}
and its diagonal part is the particle density
\begin{equation}
\rho(x,t) = \frac{1}{N}  \sum_{n=1}^N | \varphi_n(x, t) |^2    \, .
\label{density}
\end{equation}

\begin{figure}[thb]
\resizebox{3.0in}{1.9in}
{\includegraphics{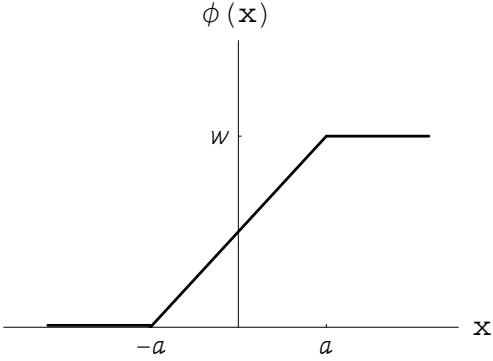}}
\caption{Phase pattern leading to the generation of a single bright-dark
quasisolitons pair.}
\label{phase}
\end{figure}

At zero temperature particles occupy the lowest possible levels. The 
corresponding eigenstates are just the plane waves
$\varphi_n(x)=\exp{(i\, p_n x /\hbar)}/\sqrt{L}$ with momenta $p_n$
quantized according to 
$p_n = 2\pi\hbar\, n /L$, where $n=0,\pm 1,\pm 2, ...$.
The evolution of the single-particle wave function can be calculated 
based on the propagator technique
\begin{equation}
\varphi_n (x,t) = \int K(x,x^{\,\prime},t) \,
\varphi_n (x^{\,\prime},0) \, dx^{\,\prime} \, ,
\label{prop}
\end{equation}
where $K(x,x^{\,\prime},t)$ is the propagator function for a one-dimensional 
box with periodic boundary conditions
\begin{eqnarray}
K(x,x^{\prime},t) = \frac{1}{L^3} \sum_n
\exp{\left[ i \left( \frac{p_n}{\hbar} (x-x^{\prime}) -
\frac{p_n^2}{2m\hbar} t \right)  \right]}   \; .
\label{propagator}
\end{eqnarray}
It is convenient now to increase simultaneously
the number of particles and the length of the box in such a way that their
ratio (i.e. the Fermi momentum) remains constant. Increasing both quantities
to infinity defines the, so called, thermodynamic limit. In this limit the 
propagator takes the same form as for a free particle and is given by
\begin{eqnarray}
K(x,x^{\prime},t) = \sqrt{\frac{m}{iht}} 
\exp{\left[ i \frac{m}{2\hbar t} (x-x^{\prime})^2 \right]}  \, .
\label{propagator1}
\end{eqnarray}

To simplify further calculations we assume the following form of the imprinted
phase (see Fig. \ref{phase})
\begin{equation}
\phi(x) =
\left\{
   \begin{array}{lll}
   0  &, &\,  x < -a \\
   \frac{w}{a} (\frac{x}{a}+1)  &, &\,   -a \leq x \leq a   \;\;,  \\
   w &, &\,  x > a  \\                   
   \end{array}
\right.
\end{equation}
where the parameters $a$ and $w$ define the width and the jump of the
imprinted phase, respectively. The Fermi momentum in one-dimensional
space, defined as $hN/(2L)$, determines the characteristic length $\lambda_F$ 
via the relation $\lambda_F=h/p_F =2L/N$. From now on we choose the Fermi 
length as a unit of length. This quantity does not change while realizing the
thermodynamic limit procedure. Simultaneously, the expression $h/\varepsilon_F$ 
gives a unit of time. Hence, according to the formula (\ref{propagator1}) the 
single particle wave function evolves as
\begin{eqnarray}
\varphi_n(x,t) = \frac{1}{\sqrt{i2Lt}}  \int_{-\infty}^{\infty}
e^{i\frac{\pi}{2t} (x-x^{\prime})^2} e^{ik_n x^{\prime}}
e^{i \phi(x^{\prime})}  dx^{\prime}  \, ,
\label{evo}
\end{eqnarray}
where the wave number $k_n=p_n/\hbar$ was introduced.

\begin{figure}[thb]
\resizebox{3.3in}{2.5in}
{\includegraphics{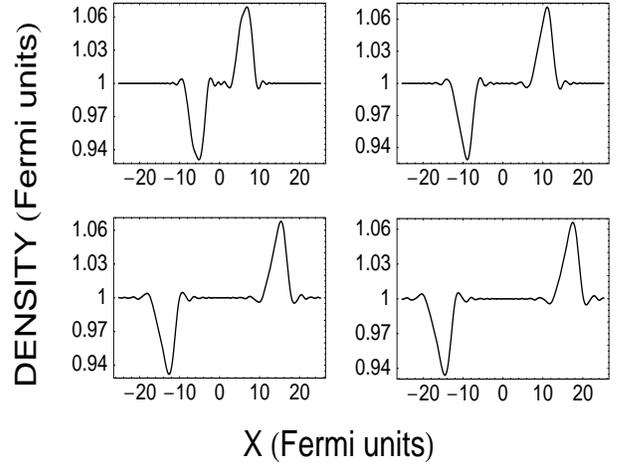}}
\caption{Evolution of a single bright-dark quasisolitons pair. Successive frames
show the fermionic density (normalized to $N/2$) of $501$ atoms
after writing a single phase step of height $\pi$ and the width
$a=4 \lambda_F$. The corresponding times are $3$, $5$, $7$,
and $8$ in units of $h/\varepsilon_F$.} 
\label{pair}
\end{figure}

Calculating the integral (\ref{evo}) one obtains an explicit formula for
the evolution of the single particle wave function 
(in units of $1/\sqrt{\lambda_F}$) 
\begin{eqnarray}
&&\varphi_n(x,t) = \frac{1}{\sqrt{i2L}}\;
e^{-ik_n^2 \frac{t}{2\pi} + ik_n x}  \left\{ (1+i) (1+e^{iw}) /2 
\right.   \nonumber  \\
&&\left. + C(u_{-a}) + i S(u_{-a}) - e^{iw} C(u_{a}) 
- e^{iw} i S(u_{a})        \right.   \nonumber  \\
&&\left. 
+ e^{i [w + \frac{w}{2a} x -  \frac{t}{2\pi}\frac{w}{a} (k_n+\frac{w}{4a})]}
\left[ C(u_{a}^{\,\prime}) + i S(u_{a}^{\,\prime}) - C(u_{-a}^{\,\prime})   
 \right.  \right.   \nonumber  \\
&&\left. \left. 
- i S(u_{-a}^{\,\prime}) \right]   \right\}  \, ,
\label{evolution}
\end{eqnarray}
where $C(z)$ and $S(z)$ are the Fresnel integrals \cite{Abram} and
their arguments are the following functions of position and time
\begin{eqnarray}
u_a & = & (a-x+k_n \frac{t}{\pi})/ \sqrt{t}   \nonumber  \\ 
u_{-a} & = & (-a-x+k_n \frac{t}{\pi})/ \sqrt{t}   \nonumber  \\ 
u_a^{\,\prime} & = & u_a + \frac{w}{2\pi a} \sqrt{t}  \nonumber  \\ 
u_{-a}^{\,\prime} & = & u_{-a} + \frac{w}{2\pi a} \sqrt{t}   \; \;.
\end{eqnarray}

The fermionic density (the diagonal part of one-particle density matrix)
is just the sum of single-particle densities obtained from single-particle
orbitals given by (\ref{evolution}). Fig. \ref{pair} shows such a density 
for a system of $501$ atoms at various times after writing a phase step of 
height $\pi$ and the width $4 \lambda_F$. As already reported in Ref.
\cite{Tomek1,Tomek2} the bright and the dark quasisolitons are generated. They 
propagate with different velocities and in opposite direction. The speed of 
bright quasisoliton is slightly higher than the speed of sound (equal to $p_F/m$)
whereas opposite is true for the dark quasisoliton. We confirm all the details
of dynamics of such a system described earlier in Ref. \cite{Tomek1,Tomek2},
especially its dependence on the ratio $a/\lambda_F$ of the width of
the imprinted phase step and the Fermi length.

\begin{figure}[thb]
\resizebox{3.0in}{1.9in}
{\includegraphics{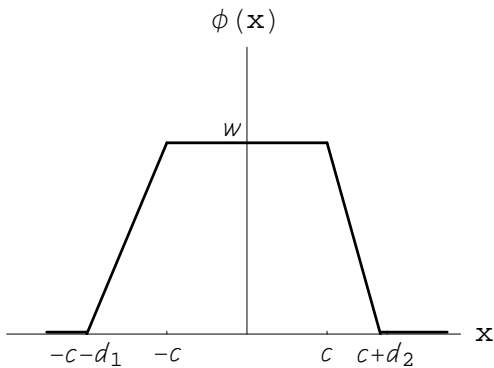}}
\caption{Phase pattern used for generating two bright-dark quasisolitons 
pairs. In this case bright quasisolitons move towards each other and collide.}
\label{phase1}
\end{figure}

The purpose of the present paper is, however, to investigate the collision 
between bright fermionic quasisolitons. To this end, we imprint the phase pattern 
as in Fig. \ref{phase1} on the fermionic gas. It results in a creation of two 
pairs of bright-dark quasisolitons with bright quasisolitons moving towards each 
other. It turns out that the time evolution of individual orbitals can be expressed 
in terms of the evolution corresponding to the phase pattern already 
considered in Fig. \ref{phase} and given by the formula (\ref{evolution}). 
First, let us notice that the phase shown in Fig. \ref{phase1} can be split 
in the following way $\phi^{lr}=\phi^l+\phi^r -w$, where $\phi^l$ and $\phi^r$
are the left and the right cuts of the phase shown in Fig. \ref{phase1}, 
respectively. Next, using the formula (\ref{evo}) one gets
\begin{eqnarray}
\varphi_n^{lr}(x,t) = \varphi_n^{l}(x,t) + \varphi_n^{r}(x,t) - A  \;,
\label{philr}
\end{eqnarray}
where
\begin{eqnarray}
A & = & \frac{1}{\sqrt{i2Lt}}  \int_{-\infty}^{\infty}
e^{i\frac{\pi}{2t} (x-x^{\prime})^2} e^{ik_n x^{\prime}}
e^{i w}  dx^{\prime}   \nonumber \\   
& = & \frac{1+i}{\sqrt{i2L}}\, 
e^{i w} e^{-ik_n^2 \frac{t}{2\pi} + ik_n x} 
\end{eqnarray}
and superscripts '$l$' and '$r$' indicate that the wave functions are 
propagated after the imprinting the single step patterns $\phi^l$ and  $\phi^r$,
respectively. The superscript '$lr$' means the evolution according to the phase 
pattern plotted in Fig. \ref{phase1}. 

\begin{figure}[thb]
\resizebox{3.3in}{2.2in}
{\includegraphics{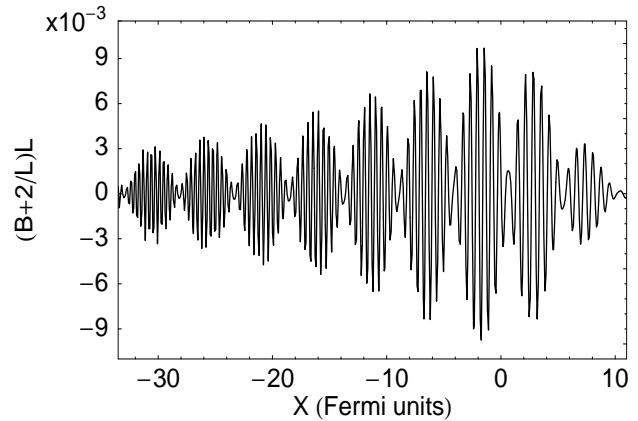}}
\caption{The relative (with respect to $1/L$) oscillations of the expression 
$B+2/L$ as a function of position for a particular $(n=200)$ single-particle 
orbital. The length of the box $L=250 \lambda_F$ and the time is $12$ in units 
of $h/\varepsilon_F$. Other parameters: $c=15, d_1=4, d_2=5,$ in units of 
$\lambda_F$, and $w=\pi$ (the case of the last frame of Fig. \ref{coll}). }
\label{diff}
\end{figure}

Both orbitals $\varphi_n^{l}(x,t)$ and $\varphi_n^{r}(x,t)$ can be
calculated based on the following rules that connect the time
evolution due to the phase pattern given in Fig. \ref{phase} with
the evolution resulting from patterns obtained by the shift 
($\varphi_n^{(s)}(x,t)$), the reflection ($\varphi_n^{(ref)}(x,t)$), 
and a combination of the reflection and the shift ($\varphi_n^{(rs)}(x,t)$) 
of that one shown in Fig. \ref{phase}
\begin{eqnarray}
\varphi_n^{(s)}(x,t) & = & e^{ik_n s} \varphi_n(x-s,t) \nonumber \\
\varphi_n^{(ref)}(x,t) & = & \varphi_{-n}(-x,t) \nonumber \\
\varphi_n^{(rs)}(x,t) & = & e^{-ik_n s} \varphi_{-n}(-x-s,t)  \; .
\end{eqnarray}
One has 
\begin{eqnarray}
\varphi_n^l(x,t) & = & e^{-ik_n (c+d_1/2)} \varphi_n \left(x+c+\frac{d_1}{2},t;
a=\frac{d_1}{2}\right)   \nonumber  \\
\varphi_n^r(x,t) & = & e^{ik_n (c+d_2/2)} \varphi_{-n} \left(-x+c+\frac{d_2}{2},t;
a=\frac{d_2}{2}\right)  \nonumber  \, .  \\
\label{philphir}
\end{eqnarray}

\begin{figure}[thb]
\resizebox{3.3in}{3.3in}
{\includegraphics{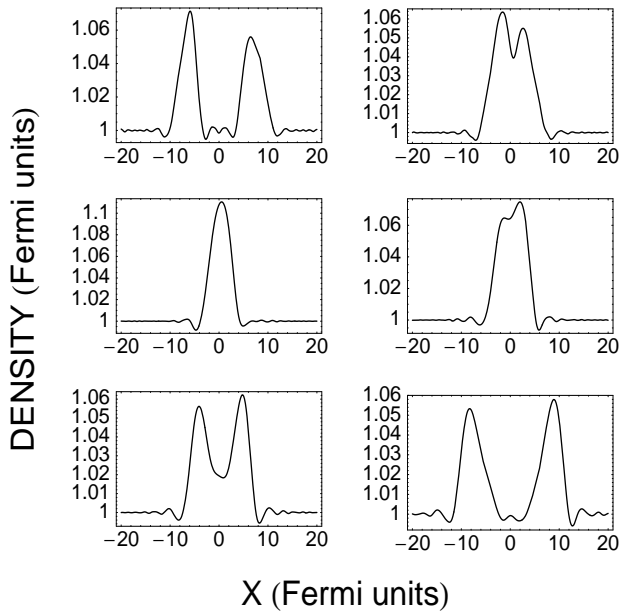}}
\caption{Two bright quasisolitons collision. Successive frames show the fermionic 
density (normalized to $N/2$) of $501$ atoms after writing two phase 
step pattern (as in Fig. \ref{phase1}) of height $\pi$ and the widths equal to
$4 \lambda_F$ (the left one) and $5 \lambda_F$ (the right one). The corresponding 
times are $5$, $7$, $8.3$, $9$, $10$, and $12$ in units of $h/\varepsilon_F$. 
Other parameters: $c=15$ in units of $\lambda_F$. }
\label{coll}
\end{figure}

The relation (\ref{philr}) can be rewritten now in terms of the single-particle
densities
\begin{eqnarray}
&& \rho_n^{lr}(x,t) = \rho_n^{l}(x,t) + \rho_n^{r}(x,t) 
+ \frac{1}{L} + B   \nonumber \\
&& B = 2 Re (\varphi_n^{l}(x,t) \varphi_n^{r*}(x,t) 
- \varphi_n^{l}(x,t) A^* - \varphi_n^{r}(x,t) A^*)  \nonumber \\
\label{den}
\end{eqnarray}
and it turns out that approximately
\begin{eqnarray}
\rho_n^{lr}(x,t) & = & \rho_n^{l}(x,t) + \rho_n^{r}(x,t) 
- \frac{1}{L}   \; .
\label{den1}
\end{eqnarray}
In fact, what is left, i.e. the expression $B+2/L$, oscillates as a function 
of position around zero with a relative (in comparison with $1/L$) amplitude 
usually of the order of $10^{-4}$ right after the quasisolitons are created. This
amplitude increases in time but remains at the low level ($10^{-2}$) even after
the collision of qusisolitons. An example of that is given in Fig. \ref{diff} where 
we consider the box of length $L=250 \lambda_F$ (it corresponds to the last frame
of Fig. \ref{coll}). Only when the imprinted phase step gets very sharp the 
relative amplitude becomes as large as $10^{-1}$. The total density is the sum of 
single-particle densities and equals
\begin{eqnarray}
\rho^{lr}(x,t) & = & \rho^{l}(x,t) + \rho^{r}(x,t) - \frac{N}{2L}   \; .
\label{dentot}
\end{eqnarray}
Here, the density is normalized to the number of particles divided by 2.
Therefore, in the thermodynamic limit, i.e. when the size of the box and the 
number of atoms increase to infinity while keeping their ratio constant, one 
gets the following rule
\begin{eqnarray}
\rho^{lr}(x,t) & = & \rho^{l}(x,t) + \rho^{r}(x,t) -1    \;.
\label{den2}
\end{eqnarray}
First, this formula generalizes the results already obtained in Ref. \cite{Tomek2}
(and expressed by solution (24)) to any perturbation of the initial density.
Secondly, it happens that going from (\ref{den1}) to (\ref{dentot}) the ''noise" 
present for each orbital (see Fig. \ref{diff}), even if it is large,
adds destructively causing that the relation (\ref{den2}) is satisfied almost
perfectly. It has to be noticed that the same property of additivity (\ref{den2}) 
is shared by the solutions of the wave equation. Here, however, single-particle
densities satisfy the nonlinear hydrodynamic equations according to the Madelung
representation of the Schr\"odinger equation \cite{Madelung}
\begin{eqnarray}
&&\frac{\partial \rho_n}{\partial t} + \frac{\partial}{\partial x} 
\left( \rho_n v_n \right) = 0        \nonumber   \\
&&\frac{\partial v_n}{\partial t} + \frac{\partial}{\partial x} 
\left( \frac{v_n^2}{2}      - \frac{\hbar^2}{2 m^2} 
\frac{1}{\sqrt{\rho_n}}  \frac{\partial^2}{\partial x^2} \sqrt{\rho_n}   
\right) = 0   \, ,  
\label{Madrep}
\end{eqnarray}
where $\rho_n$ and $v_n$ are the density and the velocity fields, respectively,
associated with the orbital $\varphi_n^{lr}(x,t)$.

In Fig. \ref{coll} we illustrate the collision of two bright fermionic 
quasisolitons. Successive frames show the total density calculated based on the 
formulas (\ref{philr}), (\ref{philphir}), and (\ref{evolution}). No interference, 
as a consequence of (\ref{den2}), is observed when both peaks meet at the center 
of the box what distinguishes this case from the case of collision of usual wave 
packets. It is not surprising, since building the quasisoliton we add probabilities 
(see (\ref{density})) rather than the amplitudes. After the collision both quasisolitons
reappear, however no phase shift, usually expected when two solitons collide, is 
observed. This is another consequence of  formula (\ref{den2}). So, what we have is a 
solution (in a sense of (\ref{density})) of a set of nonlinear hydrodynamic equations 
(\ref{Madrep}) that is a sum of quasisolitons not only well before and after the 
collision but all the time.

In conclusion, we have investigated the dynamics of fermionic quasisolitons generated
in ultracold fermionic gas by using the technique of phase imprinting. We obtained
an analytic formula that describes the propagation of a bright-dark quasisoliton pair.
Based on this formula we were able to study the collision of two bright fermionic 
qusisolitons and found that the quasisolitons are subject to the superposition
principle -- the sum of two quasisolitons is a two-quasisoliton solution.

\acknowledgments
We thank M. Gajda and K. Rz\c{a}\.{z}ewski for stimulating discussions.
E. W. was supported by the Polish KBN Grant No. 2 P03B 097 22.
M. B. acknowledges support by the Polish Ministry of Scientific
Research Grant Quantum Information and Quantum Engineering
No. PBZ-MIN-008/P03/2003.

\end{document}